\documentclass[aps,prl,showpacs,preprint,floats]{revtex4}

\usepackage{amsfonts,amssymb,graphicx}
\usepackage[dvips]{color}
\usepackage[latin1]{inputenc}

\newcommand{\nc}{\newcommand}

\nc{\be}{\begin{equation}}
\nc{\ee}{\end{equation}}
\nc{\bea}{\begin{eqnarray}}
\nc{\eea}{\end{eqnarray}}
\nc{\ba}{\begin{array}}
\nc{\ea}{\end{array}}
\nc{\ov}{\overline}
\nc{\wt}{\widetilde}

\begin{document}

\title{\mbox{Dependence of the critical temperature}\\
on the Higgs field reparametrization}

\author{E. Di Grezia}
\email{digrezia@na.infn.it} \affiliation{\small Universit\`a
Statale di Bergamo, Facolt\`a di Ingegneria, viale Marconi 5,
I-24044 Dalmine (BG), Italy \\ and Istituto Nazionale di Fisica
Nucleare, Sezione di Milano, via Celoria 16, I-20133 Milan, Italy}
\author{S. Esposito}
\email{salvatore.esposito@na.infn.it}%
\affiliation{\small Dipartimento di Scienze Fisiche,
Universit\`{a} di Napoli ``Federico II'' \\ and Istituto Nazionale
di Fisica Nucleare, Sezione di Napoli, Complesso Universitario di
Monte S. Angelo, via Cinthia, I-80126 Naples, Italy}
\author{G. Salesi}
\email{salesi@unibg.it} %
\affiliation{\small Universit\`a Statale di Bergamo, Facolt\`a di
Ingegneria, viale Marconi 5, I-24044 Dalmine (BG), Italy \\ and
Istituto Nazionale di Fisica Nucleare, Sezione di Milano, via
Celoria 16, I-20133 Milan, Italy }


\

\

\begin{abstract}
\noindent We show that, despite of the reparametrization symmetry
of the Lagrangian describing the interaction between a scalar
field and gauge vector bosons, the dynamics of the Higgs mechanism
is really affected by the representation gauge chosen for the
Higgs field. Actually, we find that, varying the parametrization
for the two degrees of freedom of the complex scalar field, we
obtain different expressions for the Higgs mass: in its turn this
entails different expressions for the critical temperatures,
ranging from zero to a maximum value, as well as different
expressions for other basic thermodynamical quantities.
\end{abstract}

\maketitle

\noindent The Higgs mechanism is the basic ingredient of many theories,
ranging from superconductivity to elementary particle physics,
where the phenomenon of spontaneous symmetry breaking (SSB) plays
a key role. In fact, the original idea by Higgs (and others)
\cite{Higgs,Bailin} to give a non-vanishing mass to gauge vector
bosons through the coupling with a scalar field $\phi$, in the
framework of the Nambu model of SSB \cite{Nambu}, was inspired by
the fundamental works by Ginzburg and Landau \cite{GL} aimed at
accounting the emergence of short-range electromagnetic
interactions, mediated by massive-like photons, in superconductors
and superfluids.

One of the topical issues in present day experimental research on
elementary particles is, indeed, just the search for the Higgs
boson at LHC \cite{LHC}, whose effective discovery will be the
keystone for the complete confirmation of the Standard Model of
electroweak interaction by Glashow, Weinberg and Salam \cite{GWS}.
Instead, in superconductivity mean-field theory (just to mention one
application in condensed matter physics), the scalar field
describes the dynamics of the Cooper pairs in the superconductors,
$\phi$ being interpreted as the wave function of the Cooper pairs
in their center-of-mass frame. For the sake of definitess, and
without loss of generality, in this paper we shall often refer to
superconductors. In the simplest version of the Ginzburg-Landau (GL) model (which will
be considered here), the Lagrangian density describing a scalar field
$\phi$ interacting with the electromagnetic field $A_\mu$ is given
by (hereafter we assume $m^2<0$)
\begin{equation} {\cal L}=-\frac{1}{4}F_{\mu \nu }F^{\mu \nu
}+\left(D_{\mu }\phi\right)^{\dagger }\left(D^{\mu }\phi\right) -
m^2\phi^\dagger\phi -\frac{\lambda}{4}(\phi^\dagger\phi)^2\,,
\label{1}
\end{equation}
where $\lambda$ is the self-interaction coupling giving the strength
of the Cooper pair binding, $F_{\mu\nu}\equiv\partial_\mu A_\nu -
\partial_\nu A_\mu$ is the electromagnetic field strength and
$D_{\mu }=\partial_{\mu}+ 2ieA_{\mu }$ \ is the covariant
derivative for the Cooper pair with electric charge $2e$. The
field $\phi$ is assumed to be a complex quantity, with {\em two}
degrees of freedom, but only {\em one} non-vanishing vacuum
expectation value (VEV), namely the condensation value of the
field below a critical temperature $T_c$. Usually, such two
degrees of freedom are represented by the real and imaginary part
of $\phi$ (see, for example, ref.\,\cite{Bailin}), but other
representations (for example, in terms of modulus and phase of
$\phi$) are of course possible. As is well known, the above
Lagrangian is invariant under reparametrization of $\phi$, and the
various representations of the complex field describe. at the
classical (tree) level, the same physical reality. This is true
also in two (or more) Higgs models, where it holds the
``rephasing'' invariance of the Higgs multiplet (see
refs.\cite{DGG}). Already in 1984, however, some authors
\cite{NXC} pointed out that different representations of $\phi$
can lead to {\em different} expressions for the critical
temperature $T_c$ where the condensation phenomenon takes place,
since the dynamics ruled by different degrees of freedom could
lead, in principle, to different predictions \cite{TwoTC}, when
considering also the radiative, non-tree corrections.
Nevertheless, it should be emphasized that such a difference is
purely formal {\em if only one condensation occurs} since, in this
case, the different dependence of $T_c$ on the model parameters
$m^2,\, \lambda$ is not observable, being such parameters not
directly measurable (they can be deduced just by measurement of
$T_c$ or other observables). The situation is, instead, just
different if two (or more) condensations take place in peculiar
physical system as, e.g., same non-standard superconductors. In
this case, the difference between the critical temperatures leads
to distinct superconductive phases, entailing two discontinuities
in the specific heat and unusual magnetic properties. In a recent
series of papers of ours \cite{TwoTC, Others, Ruth} we have,
indeed, studied in detail this possibility, and it is very
remarkable that the two different representations mentioned above
(real and imaginary parts versus modulus and phase) account, in a
very simple manner, for the apparently exotic properties observed
in the superconductivity of strontium ruthenate \cite{Ruth, Mac,
Deg}. These observations evidently urge to consider accurately the
problem of the field reparametrization in the Higgs mechanism:
actually, in the present paper we shall expound a sufficiently
general and comprehensive analysis of the physical implications of
the field representation gauge. We start our analysis from the
very general assumption that the complex scalar field $\phi(x)$
appearing in (\ref{1}) is endowed with two different degrees of
freedom described by the real scalar fields $a(x)$, $b(x)$. The
only non-vanishing VEV of $\phi(x)$ is introduced by means of the
(non-zero) real parameter $a_0$ as follow \footnote{For the sake
of comparison, we adopt the same normalization as in
Ref.\,\cite{Bailin}.}
\begin{equation}
< \! \phi(x) \! > = \frac{a_0}{\sqrt{2}}\,. \label{2}
\end{equation}
Without loss of generality we assume that such VEV corresponds
to $a=a_0$, $b=0$, that is, the field $a(x)$ condenses for
$a=a_0$, while the field $b(x)$ does not condense. Furthermore, we
also assume that the field $\phi(x)$ can be expanded in a Taylor
series around such values; on very general grounds, we can then
represent the field $\phi(x)$ with the following formula:
\begin{equation}
 \phi(x)  = \sum_{n,m} a_0^{1-n-m}(c_{nm} + i d_{nm})a^n b^m\,.
 \label{3}
\end{equation}
Here $c_{nm},\, d_{nm}$ are real, dimensionless coefficients that
characterize the different representations, while the factor
$a_0^{1-n-m}$ is introduced for dimensional reasons (both the fields
$a,b$ have the same physical dimensions as $a_0$). Notice that,
in order to preserve the appearance of {\em two} degrees of
freedom, at least one $c$-coefficient and one $d$-coefficient must
be different from zero:
\begin{eqnarray}
&& \exists n'\in \mathbb{N},\,\,\exists m'\in \mathbb{N}_0:\,\,
c_{n'm'}\neq 0, \nonumber \\ && \exists n''\in \mathbb{N}_0,
\,\,\exists m'\in \mathbb{N}:\,\, d_{n''m''}\neq 0\,. \label{4}
\end{eqnarray}
The most ``popular" parametrizations are the ``Gauss representation''
\begin{equation}
\phi =\frac{1}{\sqrt{2}}(a + ib) \label{5}
\end{equation}
where we can observe a condensation of the real (or imaginary) part;
or the ``Euler representation''
\begin{equation}
\phi =\frac{1}{\sqrt{2}}\,a\,{\rm e}^{ib/a_{0}}\,,
\label{6}
\end{equation}
where we can observe a condensation of the modulus (or phase). Correspondingly
the $c_{nm},\, d_{nm}$ coefficients take the values
\begin{eqnarray}
&& c_{nm} = \frac{1}{\sqrt{2}}\delta_{n1}\delta_{m0}
\qquad\qquad d_{nm} = \frac{1}{\sqrt{2}}\delta_{n0}\delta_{m1}
\label{7}
\end{eqnarray}
and
\begin{eqnarray}
&& c_{nm} = \frac{1}{\sqrt{2}}\frac{(-1)^k}{(2 k)!}\delta_{n1}\delta_{m,2k}
\qquad\qquad d_{nm} = \frac{1}{\sqrt{2}}\frac{(-1)^k}{(2k+1)!}\delta_{n1}\delta_{m,2k +1}\,,
\label{8}
\end{eqnarray}
respectively. In general, the coefficients $c_{nm},\, d_{nm}$ are
not completely arbitrary, but have to satisfy several constraints,
that will be discussed in the following. First of all, we take
into account that only one non-vanishing VEV exists, given by
Eq.\,(\ref{2}). By equating Eqs.\,(\ref{2}) and (\ref{3}), for
$a=a_0$, $b=0$ we obtain:
\begin{equation}
  \sum_{n} c_{n0}= \frac{1}{\sqrt{2}} \qquad\qquad \sum_{n} d_{n0}= 0\,.
 \label{9}
\end{equation}
Let us now proceed with the Higgs mechanism, by expanding $\phi$
in Eq.\,(\ref{3}) around its VEV,
\begin{equation}
a\simeq a_0 + \tilde{a}\,,\qquad\quad  b\simeq\tilde{b}
\label{10}
\end{equation}
($\tilde{a}$, $\tilde{b}$ are fluctuations fields), and
inserting the resulting expression in Lagrangian (\ref{1})
(we will consider only terms up to the second order in the fields).
The kinetic terms for the fields $\tilde{a}$ and $\tilde{b}$ coming
out from (\ref{1}) are
\begin{eqnarray}
{\cal L}_{\rm kin}&\simeq & \left[\left(\sum_{n} c_{n0}\right)^2 +
\left(\sum_{n} n d_{n0}\right)^2\right]\partial_\mu
\tilde{a}\,\partial^\mu \tilde{a} +
\left[\left(\sum_{n} c_{n1}\right)^2 + \left(\sum_{n}
d_{n1}\right)^2\right]\partial_\mu\tilde{b}\,\partial^\mu
\tilde{b} \nonumber \\ &+& 2\left[ \left(\sum_{n}
c_{n1}\right)\left(\sum_{n}n c_{n0}\right) \right.
\left. \left(\sum_{n} d_{n1}\right)\left(\sum_{n} n d_{n0}\right)\right]\partial_\mu
\tilde{a}\,\partial^\mu \tilde{b}\,.
 \label{11}
\end{eqnarray}
By diagonalizing the above Lagrangian we deduce the following constraints:
\begin{eqnarray}
&& \left(\sum_{n} c_{n0}\right)^2 + \left(\sum_{n} n
d_{n0}\right)^2= \frac{1}{2}
\nonumber \\
&& \left(\sum_{n} c_{n1}\right)^2 + \left(\sum_{n}
d_{n1}\right)^2= \frac{1}{2} \label{12}
\\
&& \left(\sum_{n} c_{n1}\right)\left(\sum_{n}n c_{n0}\right) +
\left(\sum_{n} d_{n1}\right)\left(\sum_{n} n d_{n0}\right)= 0\,.
 \nonumber
\end{eqnarray}
The potential terms for the scalar field
\begin{equation}
{\cal L}_{\rm pot}= -m^2\phi^\dagger\phi - \frac{\lambda}{4}(\phi^\dagger\phi)^2\,,
\label{13}
\end{equation}
can be analogously evaluated up to second order terms. By
taking into account the constraints in Eqs.\,(\ref{9}) and (\ref{12}),
we are able to write down the mass-potential terms in the following form
\begin{eqnarray}
{\cal L}_{\rm pot}&\simeq &
-\left[\frac{a_0^2}{2}\left(m^2+\frac{\lambda
a_0^2}{8}\right)\right] -
\left[2\frac{a_0}{\sqrt{2}}\left(m^2+\frac{\lambda
a_0^2}{4}\right)\left(\sum_{n}n c_{n0}\right)\right]\tilde{a}
\nonumber \\ &&-
\left[2\frac{a_0}{\sqrt{2}}\left(m^2+\frac{\lambda
a_0^2}{4}\right)\left(\sum_{n} c_{n1}\right)\right]\tilde{b} -
\left[\frac{2}{\sqrt{2}}\left(m^2+\frac{\lambda
a_0^2}{4}\right)\left(\sum_{n}n c_{n1}\right) \right. \nonumber \\
&&+ \lambda a_0^2 \left.\left(\sum_{n}n
c_{n0}\right)\left(\sum_{n}
c_{n1}\right)\right]\tilde{a}\tilde{b}-
\left[\left(m^2+\frac{\lambda a_0^2}{4}\right)\left(\frac{1}{2}+
\frac{1}{\sqrt{2}} \sum_{n}n(n-1) c_{n0}\right)\right.  \nonumber
\\ &&+\frac{\lambda a_0^2}{2}\left. \left(\sum_{n}n
c_{n0}\right)^2\right]\tilde{a}^2 -
\left[\left(m^2+\frac{\lambda a_0^2}{4}\right)\left(\frac{1}{2}+
\frac{1}{\sqrt{2}} \sum_{n} c_{n2}\right)\right. \nonumber \\ &&-
\frac{\lambda a_0^2}{2}\left. \left(\sum_{n}n c_{n0}\right)^2+
\frac{\lambda a_0^2}{4}\right]\tilde{b}^2\,.
 \label{14}
\end{eqnarray}
For a generic representation of the field $\phi$, the terms in
$\tilde{a}$, $\tilde{b}$  and $\tilde{a}\tilde{b}$ do not vanish,
so that, in this case, the mass eigenstates of the system are not
the fields $\tilde{a}$, $\tilde{b}$ but a linear combination of
them. However, as we will see below (see also \cite{Bailin}), the
quantity relevant for the calculation of the critical temperature,
or of other observables, is the trace of the squared mass matrix,
that is invariant under the mentioned transformation. Therefore, for
simplicity, we will limit our attention to the following part of
the potential term:
\begin{equation}
{\cal L}_{2}= -\frac{1}{2}m_a^2\tilde{a}^2
-\frac{1}{2}m_b^2\tilde{b}^2
 \label{15}
\end{equation}
with
\begin{eqnarray}
&&m_a^2=  2m^2 \left(\frac{1}{2} +
\frac{1}{\sqrt{2}}\sum_{n}n(n-1) c_{n0}\right)+ \nonumber \\ &&
~~~~~~~ + \lambda a_0^2\left[\left(\sum_{n} n c_{n0}\right)^2+
\frac{1}{2}\left(\frac{1}{2}+ \frac{1}{\sqrt{2}} \sum_{n}n(n-1)
c_{n0}\right)\right],  \label{16}
\\
&&m_b^2= 2m^2 \left(\frac{1}{2} + \frac{2}{\sqrt{2}}\sum_{n}
c_{n2}\right)+ \lambda a_0^2\left[\frac{1}{2}-\left(\sum_{n} n
c_{n0}\right)^2+\right.
\frac{1}{2}\left.\left(\frac{1}{2}+ \frac{2}{\sqrt{2}} \sum_{n}
c_{n2}\right)\right]\,. \nonumber
\end{eqnarray}
We also introduce the following quantities that will directly
enter in the expression for the critical temperature:
\begin{eqnarray}
M_a^2&\equiv & m_a^2- m_a^2(a_0=0)= \lambda
a_0^2\left[\left(\sum_{n} n c_{n0}\right)^2+ \right.
\left.\frac{1}{2}\left(\frac{1}{2}+\frac{1}{\sqrt{2}} \sum_{n}n(n-1) c_{n0}\right)\right],
\nonumber \\
M_b^2&\equiv & m_b^2- m_b^2(a_0=0)= \lambda
a_0^2\left[\frac{1}{2}-\left(\sum_{n} n c_{n0}\right)^2\right.
\frac{1}{2}\left.\left(\frac{1}{2}+ \frac{2}{\sqrt{2}}
\sum_{n} c_{n2}\right)\right]\,.
 \label{17}
\end{eqnarray}
By using the same notations as in Ref. \cite{Bailin}, the Higgs
``mass" is thus given by (of course, in $M_H^2$ the terms
proportional to $m^2$ are not included, but only the corrections
proportional to $a_0^2$ do appear)
\begin{eqnarray}
M_H^2& = & {\rm Tr}\{M_s^2\}= M_a^2 + M_b^2 =
\lambda a_0^2\left[1+ \frac{1}{2\sqrt{2}} \sum_{n}\left(n(n-1)
c_{n0}+ 2c_{n2}\right) \right]
 \label{18}
\end{eqnarray}
and, evidently, \textit{depends on the representation chosen} (through
$c_{n0}, \, c_{n2} $). Instead, from the vector boson mass term
\begin{equation}
{\cal L}_{M_V}= e^2 |\phi|^2 A_\mu A^\mu\simeq
\frac{e^2a_0^2}{2} A_\mu A^\mu\,.
\end{equation}
we obtain the photon mass from
\begin{equation}
M_V^2= e^2 a_0^2\,,
\end{equation}
that is, obviously independent of the representation chosen. In
order to get an expression for the critical temperature of the
system, we have to consider the quantum temperature-dependent,
radiative corrections to the scalar potential that, at tree level
and in its minimum (\ref{2}), is (see the first term in Eq.\,(\ref{14}))
\begin{equation}
U_0 =\frac{1}{2}m^2 a_0^2 + \frac{1}{16}\lambda a_0^4\,.
\label{21}
\end{equation}
For $e^4\ll \lambda$ we can neglect $T=0$ corrections to the
potential; by assuming also that $T^2\gg -m^2, \lambda a_0^2,
e^2a_0^2$, the $T$-dependent quantum correction term is given by
\cite{Bailin}
\begin{equation}
U_1 \simeq -\frac{4\pi^2 T^4}{90} + \frac{T^2}{24}\left[M_H^2+3
M_V^2\right]\,,
\label{22}
\end{equation}
so that the total potential can be written as
\begin{equation}
U \simeq \frac{1}{2}m^2_{eff} a_0^2 + \frac{1}{16}\lambda
a_0^4-\frac{4\pi^2 T^4}{90}\,,
\label{23}
\end{equation}
with
\begin{equation}
m^2_{\rm eff}\simeq m^2+\frac{T^2}{12 a_0^2} \left[M_H^2+3
M_V^2\right]\,.
\label{24}
\end{equation}
The critical temperature $T_c$ of the system is, then, defined by the
equation
\begin{equation}
m^2_{\rm eff}(T_c)=0\,,
\label{25}
\end{equation}
or
\begin{equation} T_c^2 = \frac{-12 m^2}{\lambda H + 3e^2}\,,
\label{26}
\end{equation}
with
\begin{equation}
H\equiv \frac{M_H^2}{\lambda a_0^2}= 1+ \frac{1}{2\sqrt{2}}
\left[\sum_{n}\left(n(n-1) c_{n0}+ 2c_{n2} \right)\right]\,.
\label{27}
\end{equation}
The key result in Eq.\,(\ref{26}) clearly shows that \textit{the observable
$T_c$ does depend on the representation chosen for the field}
$\phi$ through the $H$ term (or, what is the same, through the
Higgs mass). As already pointed out above, however, such a
dependence is only formal if the system possesses only one
critical temperature, since the parameters $\lambda$ and $m^2$
appearing in the lagrangian (\ref{1}) are not directly observable.
This is not the case, instead, for systems showing more than one
critical temperature, as extensively discussed in
\cite{TwoTC,Others}, so that it is quite relevant to discuss
the possible consequences of Eq.\,(\ref{26}).
First of all, let us observe that, for the standard
representations (\ref{5}) and (\ref{6}), we recover the known
results \cite{Bailin, NXC} namely
\begin{equation} T_c^2 = \frac{-12 m^2}{\lambda + 3e^2}\,,
\label{28}
\end{equation}
and
\begin{equation} T_c^2 = \frac{-16 m^2}{\lambda + 4e^2}\,,
\label{29}
\end{equation}
corresponding to Higgs masses $M_H^2 =\lambda a_0^2$ and $M_H^2
=\frac{3}{4}\lambda a_0^2$, respectively. It is noticeable that
the above two results (and other similar values \cite{BDK}; cf.
also \cite{NXC} for the Glashow-Weinberg-Salam model) can be here
directly obtained as particular cases of a general theory, while
in literature they come out from different and very elaborate
theoretical approaches. The latter value, for example, was derived
using the real-time Green-function approach. In general, because
of Eq.\,(\ref{27}), the $H$ term parameterizes the relative
strength between the self-interaction of the Cooper pairs (ruled
by $\lambda$) and the electromagnetic interaction (ruled by $e$).
It is quite remarkable that such parameter, and thus $T_c$,
depends on only ``two" coefficients, $c_{n0}$ and $c_{n2}$ (for
all $n$). This fact leads to relevant consequences. Firstly, since
the $d_{nm}$ coefficients do not contribute to the expression of
$T_c$ (or $M_H$), representations of $\phi$ that differ only for
the imaginary part Im$\{\phi\}$ give the same $T_c$ (and $M_H$).
However, this fact does not at all imply that we can consider just
real representations of $\phi$ (contrary to the assumption of {\em
two}, not one, degrees of freedom): in fact, from the constraints
(\ref{9}), (\ref{12}) it immediately follows that not all $d_{nm}$
coefficients can vanish. We have, then, a further limitation
since, from Eq.\,(\ref{27}), it is evident that $T_c$ (and $M_H$)
depends only on the coefficients $c_{n0},\,c_{n2}$. This means
that only the terms $a^n$ and $a^n b^2$ in the expansions of
$\phi$ contribute to $T_c$ (and $M_H$), so that representations of
$\phi$ whose real parts differ in their expansion around the VEV
($a=a_0$, $b=0$) only for odd power terms in $b$ or for $O(b^4)$
even power terms give the same $T_c$ (and $M_H$). Another
constraint on the coefficients $c_{n0},\,c_{n2}$ (which can
assume, of course, even negative values) comes from
Eq.\,(\ref{18}) by requiring that the Higgs mass squared (after
the condensation) is a positive quantity
\begin{equation}
\sum_{n}\left(n(n-1) c_{n0}+ 2c_{n2}\right)\geq -2\sqrt{2}\,.
\label{30}
\end{equation}
Provided that such condition is satisfied, from  Eq.\,(\ref{26}) we
obtain the important result that, changing the representation of
$\phi$, the critical temperature of the system cannot assume any
arbitrarily large value, but is bounded in the interval
\begin{equation}
0\leq T_c \leq T_c^{\rm max}\,,
\label{31}
\end{equation}
with
\begin{equation}
T_c^{\rm max} = 2\sqrt{\frac{-m^2}{e^2}}\,,
\label{32}
\end{equation}
corresponding to the (reverse of the) interval $M_H \in
[0,\infty)$. Eq.\,(\ref{26}) can, then, be rewritten in the
expressive form:
\begin{equation}
T_c = T_c^{\rm max}\sqrt{\frac{1}{1+ \frac{\lambda H}{3 e^2}}}\,,
\label{33}
\end{equation}
and the dependence of $T_c$ on the ``representation factor" $H$ is
depicted in Fig.\,1.
\begin{figure}
\centering
\includegraphics[scale=0.8]{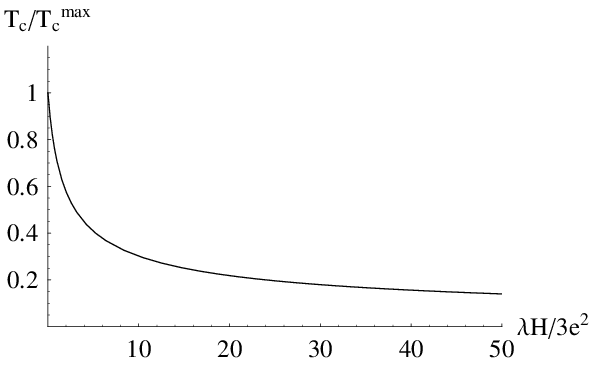}
\caption{Critical temperature versus the ``representation factor''
$H$.} \label{fig2}
\end{figure}

Note that $T_c\rightarrow 0$ (that is, no superconductivity) for
$M_H\rightarrow \infty$ ($H\rightarrow \infty$) or
$\lambda/e^2\rightarrow \infty$, that is when the Cooper
pair self-interaction is much stronger than the electromagnetic
interaction among electrons. Instead, the critical temperature
approaches its maximum value in (\ref{32}) when the Higgs mass
tends to zero ($H\rightarrow 0$) or
$\lambda/e^2\rightarrow 0$, that is for a Cooper pair
self-interaction much weaker in comparison to the electromagnetic
interaction. A possible example of representation of $\phi$ that,
independently of the value of $\lambda/e^2$, realizes such a case
(that is, for which $M_H=0$) is the following:
\begin{equation}
\phi =\frac{a}{\sqrt{2}}\left(3- 2\frac{a}{a_0}\right)+
i\frac{b}{\sqrt{2}}\left(1- 2\frac{b}{a_0}\right) \label{34}
\end{equation}
[but, of course, many other representations are possible,
according to what discussed above, provided that the constraints
(\ref{9}), (\ref{12}) and (\ref{26}) are satisfied]. Interestingly
enough, the maximum critical temperature in (\ref{32}) has an
impressive physical interpretation in terms of the entropy of the
system. In fact, from the standard expression of the free energy
$F$ in the GL model (evaluated in the VEV of the Higgs field) \cite{Others},
\begin{equation}
F =F_0 + \frac{1}{2}a(T)a_0^2 + \frac{\lambda}{16}a_0^4
\label{35}
\end{equation}
 with
\begin{equation}
a(T)=-m^2\left(1-\frac{T^2}{T_c^{2}}\right)\,,
\label{36}
\end{equation}
the entropy $S$ of the system can be easily computed from
\begin{equation}
S=-\frac{\partial F}{\partial T}\,,
\label{37}
\end{equation}
obtaining:
\begin{equation}
S=\frac{a_0^2}{12}(\lambda H + 3e^2)T = \left(\frac{M_H^2}{12}+
\frac{M_V^2}{4}\right)T\,. \label{38}
\end{equation}

\begin{figure}
\centering
\includegraphics[scale=0.8]{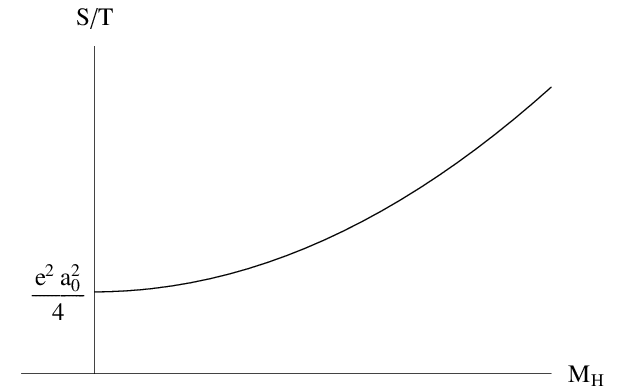}
\caption{Entropy versus the Higgs mass, for a given temperature.}
\label{fig2}
\end{figure}

\noindent For a given temperature $T$ (see Fig.2), the entropy increases for
increasing Higgs mass, starting from a minimum for $M_H=0$. Thus,
the maximum critical temperature in (\ref{32}) corresponds, for
given temperature $T$, to the minimum of the entropy of the system
(different from zero for a non-vanishing v.e.v of $\phi$) or, what
is the same, to the maximum possible order of the system. This is what
expected, since an higher temperature corresponds to a smaller Higgs mass,
which in its turn advantages the transition to the more ordered broken phase.

Summing up, we have discussed how different reparametrizations of
the scalar field ruling the Higgs mechanism (with two degrees of
freedom and one non-vanishing VEV), as described by
Eq.\,(\ref{3}), affect the expression of the critical temperature
of the system or, through the free energy (\ref{35}), all the
thermodynamical quantities of the standard GL models (applied,
e.g., to superconductors and superfluids). This study is relevant
only for physical systems that exhibit more than one critical
temperatures, as the case, for example, of the superconductivity
of strontium ruthenate \cite{Ruth}. Changing the possible
representation of the scalar fields $\phi$ results (with some
interesting exceptions, discussed above in detail) in different
values for the Higgs mass and, through this parameter, in
different critical temperatures. However, while the Higgs mass is,
in general, not bounded (ranging from $0$ to infinity), the
critical temperatures of the system can increase from zero (that
is no superconductivity) up to a maximum value $T_c^{\rm max}$,
corresponding to the non-zero minimum of the entropy system (for
given temperature). One possible representation for $\phi$
realizing such a limiting case is shown in Eq.\,(\ref{34}), and
can be regarded as a generalization of the standard representation
(\ref{5}) when highest order terms in $a,\, b$ are included.

Although we have devoted our attention namely to the standard GL
model which usually applies in condensed matter and solid state
physics, nevertheless we expect that the consequences of our study
will be not limited to that area of physics. In particular,
further studies on the elementary particle physics sector, with
special reference to the unification of the fundamental forces and
to the phase transitions in the Early Universe, can disclose
interesting new phenomena affecting our understanding of the
cosmological evolution.
 \vspace*{0.2cm}

\noindent \

\

\end{document}